# Designing defect-based qubit candidates in wide-gap binary semiconductors for solid-state quantum technologies

Hosung Seo[1,2,3], He Ma[2,4], Marco Govoni[1,2], and Giulia Galli[1,2]*

1. Materials Science Division, Argonne National Laboratory, Argonne, IL 60439, USA
2. The Institute for Molecular Engineering, The University of Chicago, Chicago, IL 60615, USA
3. Department of Physics, Ajou University, Suwon, Republic of Korea
4. Chemistry Department, The University of Chicago, Chicago, IL 60615, USA

*Corresponding author: gagalli@uchicago.edu

The development of novel quantum bits is key to extend the scope of solid-state quantum information science and technology. Using first-principles calculations, we propose that large metal ion - vacancy complexes are promising qubit candidates in two binary crystals: $4H$-SiC and $w$-AlN. In particular, we found that the formation of neutral Hf- and Zr-vacancy complexes is energetically favorable in both solids; these defects have spin-triplet ground states, with electronic structures similar to those of the diamond NV center and the SiC di-vacancy. Interestingly, they exhibit different spin-strain coupling characteristics, and the nature of heavy metal ions may allow for easy defect implantation in desired lattice locations and ensure stability against defect diffusion. In order to support future experimental identification of the proposed defects, we report predictions of their optical zero-phonon line, zero-field splitting and hyperfine parameters. The defect design concept identified here may be generalized to other binary semiconductors to facilitate the exploration of new solid-state qubits.

## I. INTRODUCTION

Optically active spin defects in wide-gap semiconductors are important resources for solid-state quantum technologies[1-4]. One well-known spin defect is the nitrogen-vacancy (NV) center in diamond[5], which may be used for applications ranging from quantum information processing[6] to quantum sensing [7,8]. Recently, alternative defect qubits in wide-gap binary semiconductors have been proposed[9-12]. In particular, di-vacancies in SiC were shown to have several desired properties similar to the diamond NV center[9,13,14] and to exhibit a quantum coherence time much longer than that of the diamond NV[13,15,16]. In a previous study, we showed that the binary nature of SiC is responsible for the improved coherence time[15]. Given the attractive properties of SiC - i.e., much cheaper than diamond and with well-established synthesis procedures - and the promising properties of its point defects, it is interesting to explore whether additional defects may be engineered in SiC as qubit candidates[17,18].

In recent years, first-principles calculations have played a key role in the search of defect qubits in wide-gap semiconductors. For example, by using density functional theory (DFT), Gali pointed out similarities between the divacancy spin in SiC and the diamond NV center[19], originating from the same $C_{3v}$ configuration of C $2sp^3$ dangling bonds in the two materials[20]. An experimental investigation of the divacancy by Koehl *et al.* readily followed[9]. Weber *et al.* formulated criteria for the systematic identification of qubits in wide-gap semiconductors and proposed to realize 'NV centers' in SiC[17], and experimental searches are underway[21,22]. First-principles DFT calculations have also been used to investigate Si vacancies ($V_{Si}$) in SiC and to identify the role of C $2sp^3$ dangling bonds in determining the properties of the optically addressable solid-state qubit[23].

The realization of 'NV-like' qubits in SiC, based on C $2sp^3$ dangling bonds, may lead to several advantageous properties[3,17,18], nevertheless a number of drawbacks are present. For example, the SiC divacancy, similar to the diamond NV center, may exhibit low optical read-out fidelity[5,24] and small ground-state spin-transverse strain coupling[25-28], which is unfavorable for certain hybrid quantum applications[29-32]. In addition, the implementation of spin qubits using C $2sp^3$ dangling bonds is not generalizable to other binary materials, e.g. nitrides. In the case of nitrides, theoretical studies have suggested that defects based on N $2sp^3$ dangling bonds, e.g. $V_{Al}O_N$ may be potential qubit candidates[33]. However, in a previous study on AlN[34], we showed that the occupied spin-orbitals of $V_{Al}O_N$ are in strong resonance with the valence band of the host, which make them unfavorable for spin qubit applications.

Therefore, it is desirable to explore the possibility of realizing qubits that are based on novel defects rather than on C or N $2sp^3$ dangling bonds. Recent theoretical studies have proposed spin defects in SiC and AlN based on cationic dangling bonds, e.g. Al $3sp^3$ states and Si $3sp^3$ states[34-36]. In a previous work, we showed that the negatively charged N vacancy in *w*-AlN could have an optically addressable spin-triplet state under a uniaxial or biaxial strain[34]. Varley *et al.* considered impurity-vacancy complexes in *w*-AlN based on Group-IV elements including Ge, Sn, Ti, and Zr[36]. They suggested that Zr- and Ti-vacancy complexes would be good candidates for spin qubits in *w*-AlN. In the case of 4*H*-SiC, Szasz *et al.* proposed that the S=1 state of the carbon-antisite vacancy defect may be stable, and hence may be a valuable qubit[35].

Using a combination of first-principles calculations, here we propose that large metal ion - vacancy (LMI-vacancy) complexes are promising qubit candidates in both 4*H*-SiC and *w*-AlN. In particular, we selected Y, La, Zr, and Hf ions for two reasons: (i) They have ionic radii larger than those of Si and Al[37], and hence they may favorably pair with anion vacancies, i.e. N vacancies in *w*-AlN and C vacancies in 4*H*-SiC. Such pairing was previously investigated for Nb in SiC[38] and Ce in AlN[39]. (ii) The selected LMIs electronegativities[40] are lower than those of Al (1.6), Si (1.9), possibly leading to the stabilization of desired charge states for the defect complexes. We found that neutral Hf- and Zr-vacancy complexes are promising candidates for spin qubits in both 4*H*-SiC and *w*-AlN. Our calculations showed that these complexes are energetically stable and exhibit a spin-triplet ground state localized in the band gap of SiC and AlN, which could be optically addressable. In addition, we predicted the optical zero-phonon line, spin zero-field splitting, and hyperfine coupling parameters of the defects, to assist future experimental detection.

The rest of the paper is organized as follow. In section II, we describe the first-principles computational methods used in this work. Our main results are presented in section III. In section IV, we discuss the unique features of the defects proposed here as potential qubits in 4*H*-SiC and *w*-AlN and we summarize our results.

## II. COMPUTATIONAL METHODS

### A. Density functional theory and $G_0W_0$ calculations

We performed DFT calculations with semi-local and hybrid functionals using plane-wave basis sets (with an energy cutoff of 75 Ry), optimized norm-conserving Vanderbilt (ONCV) pseudopotentials[41,42] and the Quantum Espresso code[43]. We used the Perdew–Burke–Ernzerhof (PBE) semi-local functional[44] and the dielectric-dependent hybrid (DDH) functional proposed in Ref.[45] with the self-consistent Hartree-Fock mixing parameter ($\alpha$) determined in Ref.[45] for SiC ($\alpha_{SiC}$= 0.15 = $1/\epsilon_{\infty,SiC}$, where $\epsilon_{\infty,SiC}$=6.5 was self-consistently computed by including the full response of the electronic density to the perturbing external electric field). For AlN, we used the PBE0 hybrid functional[46], whose choice for AlN was extensively verified in Ref.[34] (For PBE0, $\alpha_{AlN}$= 0.25, close to the self-consistently determined mixing parameter; $1/\epsilon_{\infty,AlN}$= 1/4.16 = 0.24[45]). Bulk properties of 4*H*-SiC (see Table 1) and *w*-AlN (reported in our previous study[34]) computed with the DDH functionals were found to be in excellent agreement with experimental data[40,47]. In addition, we also performed calculations with the Heyd-Scuseria-Ernzerhof (HSE06) range-separated hybrid functional[48] and projector-augmented-wave (PAW) pseudopotentials[49] to cross-check some of our results obtained with the DDH functional.

The calculation of the defect formation energy[50] of charged point defects in a crystal was carried out with the charge correction scheme developed by Freysoldt, Neugebauer, and Van de Walle[51]. We employed supercells with 480 atoms and 96 atoms for PBE and DDH calculations, respectively, and we sampled the Brillouin zone with the Gamma point only for the largest supercell and with a 2×2×2 k-point for the smallest one. Convergence studies as a function of cell size and k-meshes were reported in a previous paper [34].

The zero-phonon line (ZPL) of the LMI-vacancy complexes was obtained by calculating total energy differences (ΔSCF method) with 480-atom supercells with the PBE semi-local functional and 240-atom supercells with the hybrid functionals (DDH and HSE06). We found that energy differences computed with 480- and 240-atom supercells at the PBE level differed by less than 50 meV.

We also calculated defect level diagrams of the LMI-vacancy complexes in 4$H$-SiC and $w$-AlN within the $G_0W_0$@PBE approximation[52,53] using the WEST code[54] with 240-atom supercells and the Γ point only (see Fig. S1 for convergence tests). Table 2 compares the band-gap of diamond, 4$H$-SiC, and $w$-AlN obtained with the $G_0W_0$@PBE as well as with hybrid DFT calculations, showing excellent agreement with experiment.

### B. Spin Hamiltonian: zero-field splitting and hyperfine parameters

The properties of a defect in a crystal with spin S > 1/2, interacting with a nuclear spin $I$ can be described by the following spin Hamiltonian[55]:

$$H = \vec{S}^T \cdot \overleftrightarrow{D} \cdot \vec{S} + \vec{S}^T \cdot \overleftrightarrow{A} \cdot \vec{I}, \tag{1}$$

where $\overleftrightarrow{D}$ is the zero-field splitting (ZFS) tensor describing the splitting and the mixing of levels with different values of magnetic spin quantum number (e.g. $m_s = 0, \pm1$ for $S=1$), occuring even in the absence of an applied magnetic field and $\overleftrightarrow{A}$ is the hyperfine tensor describing the coupling between the electron spin and the nuclear spin. The first term of Eq. 1 can be written as:

$$H_{ZFS} = D_{xx}S_x^2 + D_{yy}S_y^2 + D_{zz}S_z^2 = D\left(S_z^2 - \frac{S(S+1)}{3}\right) + E(S_x^2 - S_y^2), \tag{2}$$

where $D = 3D_{zz}/2$ and $E = (D_{xx} - D_{yy})/2$ are called the axial and rhombic ZFS parameters, respectively, and the ZFS tensor $\mathbf{D}$ is traceless[55]. Hence, in the case of spin $S=1$, the $D$ term describes the energy splitting between the $m_s=\pm1$ and $m_s=0$ spin sub-levels, while the $E$ term mixes the spin sub-levels. In the case of C$_{3v}$ symmetry, the $E$ term is zero and the C$_{3v}$ axis aligns with the spin quantization axis of the $\mathbf{D}$ tensor.

For a defect spin in a crystal composed of light elements (such as Si and C), the interactions contributing to the ZFS tensor are known to be dominated by the magnetic dipole-dipole interaction between the constituent electron spins ($H_{dd}$)[56]. For instance, for a defect system with $S=1$ composed of only two unpaired electrons ($s_1=1/2$, $s_2=1/2$, and $S=s_1+s_2$), the general form of the magnetic dipole-dipole coupling is given by:

$$H_{dd} = \frac{\mu_0}{4\pi} \frac{(\gamma_e \hbar)^2}{|\vec{r}_1 - \vec{r}_2|^5} \left( r^2 \vec{s}_1 \cdot \vec{s}_2 - 3(\vec{s}_1 \cdot (\vec{r}_1 - \vec{r}_2))(\vec{s}_2 \cdot (\vec{r}_1 - \vec{r}_2)) \right), \tag{3}$$

where $\mu_0$ is the vacuum magnetic permeability, $\gamma_e$ is the electron gyromagnetic ratio, $\hbar$ is the Planck constant divided by $2\pi$, $\vec{s}_1$ and $\vec{s}_2$ are the spin-1/2 operators for the two electrons, $\vec{r}_1$ and $\vec{r}_2$ are the positions of the electrons, and $r$ is the distance between them. Using the total spin ($\vec{S} = \vec{s}_1 + \vec{s}_2$) and averaging over the spatial coordinates, one can derive an expression for the ZFS tensor's components originating from the magnetic dipole-dipole interaction of Eq. 3:

$$D_{ab} = \frac{1}{2} \frac{\mu_0}{4\pi} (\gamma_e \hbar)^2 \left\langle \Psi_{ij}(\vec{r}_1, \vec{r}_2) \left| \frac{r^2 \delta_{ab} - 3 r_a r_b}{r^5} \right| \Psi_{ij}(\vec{r}_1, \vec{r}_2) \right\rangle, \tag{4}$$

where $a$ and $b$ label the Cartesian coordinates and $\Psi_{ij}(\vec{r}_1, \vec{r}_2)$ is the wavefunction of the two-electron system.

For many-electron systems such as the LMI-vacancy spins considered here, we computed the D-tensor's components following Ref.[57]:

$$D_{ab} = \frac{1}{2} \frac{\mu_0}{4\pi} (\gamma_e \hbar)^2 \frac{1}{S(2S-1)} \sum_{i>j}^{occupied} \chi_{ij} \left\langle \Psi_{ij}(\vec{r}_1, \vec{r}_2) \left| \frac{r^2 \delta_{ab} - 3 r_a r_b}{r^5} \right| \Psi_{ij}(\vec{r}_1, \vec{r}_2) \right\rangle \tag{5}$$

where $\Psi_{ij}(\vec{r}_1, \vec{r}_2)$ is a Slater-determinant approximated by using the $i$-th and $j$-th Kohn-Sham wavefunctions of a given spin defect. The sum in Eq. (5) is over all the possible pairs of occupied Kohn-Sham wavefunctions. $\chi_{ij}$ is +1 (-1) for parallel (antiparallel) spins. As suggested in Ref.[57], we computed Eq. (5) in Fourier space; we used PBE wavefunctions obtained with a 480-atom supercell with the $\Gamma$ point only.

Our results for diamond and SiC, obtained with the ONCV norm-conserving pseudopotentials (see Table 3) systematically overestimate the experimental ZFS parameters by 200 ~ 300 MHz [58,59]. The possible numerical origin of the discrepancy, including the choice of the pseudopotential, are discussed in the Supplementary Information (SI).

The hyperfine parameters were calculated by first obtaining the ground-state wavefunctions of a LMI-vacancy spin at the PBE level of theory, with the PAW pseudopotentials, and the 480-atom supercell (Gamma-only calculations). We then calculated the hyperfine parameters by using the gauge-including projector-augmented wave method[60] (GIPAW) as implemented in the GIPAW module of the Quantum Espresso code. The core polarization effects[61] were included throughout all of our calculations.

## III. RESULTS

### A. Electronic properties of metal ion-vacancy complexes

As a validation step of the computational strategy applied to LMI-vacancy complexes, we first applied the DDH hybrid functionals to the diamond NV center and a divacancy defect (the (*hh*)-divacancy) in 4*H*-SiC, which have the same $C_{3v}$ symmetry as that of the complexes studied here; we compared our results with those already present in the literature[28,62,63] and found good agreement (see Fig. S2).

We then computed the atomic and electronic structure of the Hf- and Zr-vacancy in 4*H*-SiC using the DDH functionals. Fig. 1a shows the structure of a Hf-vacancy defect complex in 4*H*-SiC in a neutral charge state. Our hybrid functional calculation showed that substitutional Hf does not occupy the original Si site, rather it is significantly off-centered (by 0.41 A), closer to the C vacancy site, which provides extra space to accommodate the large substitutional Hf. As noted earlier, the electronegativity of this Hf (1.3[40]) is smaller than that of Si (1.9), indicating that substitutional Hf would transfer four valence electrons to the nearest neighboring C and Si dangling bonds, thus remaining in a 4+ oxidation state. Therefore, the defect geometry includes three passivated C $sp^3$ dangling bonds around substitutional Hf, and three Si $3sp^3$ dangling bonds in the $C_{3v}$ symmetry, with one $e^-$ from each Si dangling bond and one $e^-$ transferred from Hf.

Fig. 1b and 1c show the defect level diagram of the neutral Hf-vacancy complex in 4*H*-SiC and its spin density, respectively, with a fully occupied *a* state and two degenerate $e_x$ and $e_y$ states with two unpaired electrons localized within the band gap of the crystal. Although there are significant contributions from Hf and the nearby C atoms to the defect spin density, the major contribution arises from the Si $3sp^3$ dangling bonds. Hence, one may qualitatively understand the level diagram of Fig. 1b, as originating from a $C_{3v}$ configuration of three Si dangling bonds with four electrons, corresponding to a $^3A_2$ spin-triplet state, analogous to that of the diamond NV or the SiC (*hh*)-divacancy. A spin-conserving intra-defect optical excitation would then be allowed, by promoting an *a* electron to the *e* manifold in the spin-down channel, leading to a $^3E$ excited state[62]. We also found that the Zr-vacancy showed very similar properties in terms of geometrical and electronic structures (Zr belongs to the same row of the periodic table as Hf).

The energy levels of the occupied and unoccupied doubly degenerate *e* states of these defects were also computed with the $G_0W_0$@PBE method and the HSE06 functionals (see Table 4) for validation purposes. We found that all three methods yielded consistent results for the position of the levels, which are calculated to be about 1 eV above the valence band edge in SiC.

We note that the same type of defect may also be considered for optically addressable spin qubits in *w*-AlN as the electronegativities[40] of Hf (1.3) and Zr (1.3) are smaller than those of Al (1.6) and N (3.0) and their ionic radii are larger than that of Al[37]. Fig. 2a shows the defect level diagram of a Hf-vacancy complex in *w*-AlN, in which substitutional Hf is paired with a N vacancy along the [0001] direction. The metal ion passivates the N 2sp$^3$ dangling bonds and transfers one electron to the nearest neighboring Al 3sp$^3$ dangling bonds in the C$_{3v}$ configuration. The defect level diagram is qualitatively the same as that of the Hf-vacancy in 4$H$-SiC. Using the G$_0$W$_0$@PBE method and hybrid functionals, we calculated the energy levels of the occupied $e$ states to be about 3 eV below the conduction band edge (See Table 4). As shown in Fig. 2b, the dominant contribution to the ground-state spin density originates from the Al 3sp$^3$ dangling bonds, but there are also significant contributions from substitutional Hf and the nearby N atoms.

Similar defect complexes may be obtained with other LMIs, for example, La-vacancy and Y-vacancy complexes. La and Y have large ionic radii[37] and small electronegativities[40], but only three valence electrons. Hence, they may behave similar to the neutral Hf-vacancy when negatively charged. The defect level diagrams of the negatively charged La-vacancy and Y-vacancy complexes in 4$H$-SiC and $w$-AlN are reported in Fig. S3 and S4, respectively, showing, as expected, the presence of localized $e$ states similar to Fig. 1b and 2a.

We now turn to discuss the energetic stability of LMI-vacancy complexes in 4$H$-SiC and $w$-AlN.

### B. Defect stability

We investigated the stability of the LMI-vacancy defects by (1) examining the stability of the C$_{3v}$ S=1 high-spin state against potential symmetry-lowering structural distortions; and (2) investigating defect formation energies as a function of charge states. We then computed the charge transition levels and the ionization energies of the defects, which we compared to their optical zero-phonon lines (ZPLs).

In Table S1, we report the total energy differences between the S=0 singlet state (C$_{1h}$ structure) and the S=1 state (C$_{3v}$ structure) of the LMI-vacancy defects in $w$-AlN and 4$H$-SiC calculated using the DDH-DFT. We found that in all cases, the S=1 state is lower in energy than the S=0 state, e.g. by 205 (380) meV for the Hf-vacancy in 4$H$-SiC ($w$-AlN). In addition, we tested the stability of the defect geometry against perturbation to the metal ion position, to investigate whether other low-energy configurations of the defect were accessible, with small or no energy barriers, close to the proposed S=1 state. We considered in- and out-of-plane displacements of the metal ion: the former would lower the defect symmetry while the latter would lead to a different electronic structure due to a different interaction between the metal ion and the Si or Al dangling bonds. We found that the C$_{3v}$ structure shown in Fig. 1a is the lowest energy minimum structure of the defects in 4$H$-SiC and $w$-AlN at T=0 K, indicating the robustness of the S=1 state against structural distortions.

Next, we examined additional charge states. Fig. 3a and 3b show the defect formation energy of the LMI-vacancy complexes (Hf and Zr, and La, respectively) in 4$H$-SiC in the C-poor limit. The results for the C-rich case and those of Y-related defects are reported in Fig. S5 and S6.

In all cases, we found that the formation energy of a LMI-vacancy complex is lower than the sum of the formation energies of an isolated LMI impurity and an isolated C vacancy across the entire Fermi level range, regardless of the charge state. As shown in Fig. 3a (Hf and Zr in SiC), the energy gain by forming a LMI-vacancy complex is ~1 eV near the valence band maximum (VBM), and larger than ~2 eV near the conduction band minimum (CBM). For the La case, the energy gain is larger than for the Hf-vacancy: ~2 eV and ~ 3 eV near the VBM and CBM, respectively. The energy differences are the same in the C-rich limit as shown in Fig. S5. In addition, we found that the LMI-vacancy defect formation energies are lower than that of the divacancy, which was shown to be a stable defect in SiC[64]. This result strongly supports our hypothesis that the pairing of large metal ions with C vacancies leads to the formation of stable complexes in SiC.

The results of Fig. 3 also show the relative stability of different charge states. We recall that the slope of the defect formation energy as a function of the Fermi level represents the charge state of a given defect: a neutral state and a negative state are stable in a Fermi level range where the defect formation energy with slope of 0 and -1, respectively, has the lowest energy. In particular, Fig. 3a shows that the neutral Hf- and Zr-vacancy complexes with S=1 are stable in the mid-gap region of 4$H$-SiC, with (+1/0) charge transition levels (CTLs) of 1.84 eV and 1.87 eV, respectively, with respect to the CBM. This indicates that the neutral Hf-vacancy and Zr-vacancy complexes may exist in highly insulating 4$H$-SiC crystals. The negatively charged state of the La- and Y-vacancy, with S=1 is stable near the conduction band edge with the (0/-1) CTLs of 0.86 eV and 0.99 eV, respectively.

Our results for the formation energies of the LMI-complexes in $w$-AlN are similar to those for SiC, as shown in Fig. 4. The Hf- and the La-vacancy are stable in neutral and negatively charged states, and the formation energy of the Zr-vacancy is similar to that of the Hf-vacancy. The (+1/0) CTL of the Hf- and the Zr-vacancy are 2.76 eV and 2.84 eV, respectively, with respect to the CBM. The stability region for the neutral Hf- and Zr-vacancies is shown in Fig. 4a as a grey shaded area, and it overlaps with that of the neutral N vacancy, which has been previously detected in experiment[65]. Furthermore, the defect formation energy of the neutral Hf-vacancy is smaller than the sum of an isolated Hf impurity and an isolated N vacancy formation energies, indicating that realizing the S=1 state of the Hf-vacancy complex is indeed possible. The same conclusion was obtained for the Zr-vacancy. The negative charge state of the La-vacancy is stable near the CBM, with the (0/-1) CTL position 1.43 eV below the CBM. We also found a significant energy gain (1~2 eV) upon formation of the La-vacancy complex from an isolated La impurity and an isolated N vacancy across the entire band gap.

### C. Zero-phonon lines of the LMI-vacancy complexes

The optical initialization and readout of the diamond NV center and the SiC divacancy relies on the spin-conserving excitation to a $^3E$ spin-triplet excited state and its spin-selective decay[9,58]. We found that the same spin-conserving excitation scheme may occur in the LMI-vacancy complexes in 4$H$-SiC and $w$-AlN, as shown in Table 5, where we report calculated ZPLs using total energy differences (ΔSCF calculations) at the PBE, the DDH, and the HSE06 levels of theory. We note that the DDH and HSE06 calculations yielded similar results.

The calculated ZPLs are 1.7 eV (PBE) and 2.2 eV (Hybrids) for the diamond NV center and 1.0 eV (PBE) and 1.3 eV (Hybrids) for the SiC ($hh$)-divacancy; our PBE results underestimate the experimental ZPLs (1.945 eV and 1.094 eV) and our hybrid functional results consistently overestimate them by 0.2~0.3 eV. We computed the ZPLs of the Hf-vacancy and Zr-vacancy complexes in 4$H$-SiC to be ~2.0 eV using the DDH and HSE06 functionals. These calculations were not conducted at the PBE level of theory as the occupied $a$ state is deep in the valence band due to the PBE band gap underestimation. We expect our hybrid functional results to provide an upper bound to the measured ZPLs of the Hf- and Zr-vacancy in 4$H$-SiC, similar to our diamond NV and SiC divacancy results; we would estimate the measured ZPLs to be close to ~1.7 eV. Similarly, we suggest that the measured ZPLs of the Hf- and Zr-vacancy in $w$-AlN are between ~2.3 eV (PBE, lower bound) and ~3.0 eV (hybrid, upper bound).

For the negatively charged La-vacancy, the corresponding computed ZPLs are 1.20 (1.57) eV and 2.24 (2.82) eV in 4$H$-SiC and $w$-AlN, at the PBE (DDH) level of theory. However, the (0/-1) CTLs of the La-vacancy in 4$H$-SiC and $w$-AlN were found to be 0.86 eV and 1.43 eV, respectively, with respect to the CBM. This indicates that the $^3E$ excited state of the negatively charged La-vacancy is above the conduction band edge in both 4$H$-SiC and $w$-AlN, which may lead to the ionization of the defect center. This turned to be also the case for the negatively charged Y-vacancy as its (0/-1) CTL is very shallow (See Fig. S6). Therefore, in what follows we do not further consider the negatively charged La-vacancy and Y-vacancy complexes, and focus on the Hf-vacancy and the Zr-vacancy complexes for use as potential qubits in 4$H$-SiC and $w$-AlN.

D. **Spin Hamiltonian Parameters: Zero-field splitting and hyperfine interaction**

Electron paramagnetic resonance (EPR) is a powerful technique to detect and characterize paramagnetic defects in solids[55]. The zero-field splitting D tensor and the hyperfine A tensor are key components of the spin Hamiltonian that determines the EPR spectrum (see Eq. 5). For the Hf-vacancy and Zr-vacancy in SiC (AlN), we found $D$ = 1.40 (2.96) GHz and 1.10 (3.05) GHz, respectively, using the ONCV pseudopotentials, as reported in Table 3. These values are comparable to those of the diamond NV and the SiC divacancy, which were measured to be 2.9 GHz[58] and 1.3 GHz[59], respectively.

In order to study the coupling between defect spin qubits and lattice strain, we computed $D$ as a function of hydrostatic pressure, $D(P)$[66]. In particular, we investigated the role of different dangling bonds (e.g. C $2sp^3$ vs. Si $3sp^3$) and different type of host crystals (e.g. diamond vs. SiC or AlN) in determining the coupling characteristic of spin to strain. We considered hydrostatic pressure, which may yield an isotropic compressive strain around the defect centers, thus preserving the $C_{3v}$ symmetry. Defect qubits under hydrostatic pressure could also be easily accessible in diamond anvil cell experiments[66]. We first compare $D(P)$ of the diamond NV and the SiC divacancy, and then discuss $D(P)$ of the LMI-vacancy complexes.

Fig. 5a shows that in diamond, $D(P)$ is linear up to 100 GPa, while in SiC, $D(P)$ deviates from a linear behavior already at 50 GPa (SiC is known to be stable under pressure up to 100 GPa[67]). The linear behavior found for the diamond NV is in good agreement with previous experimental[66] and theoretical[68] results. We found a slope of 10.91 MHz/GPa, compared to an experimental value of 14.58 MHz/GPa[66] and a previous theoretical value of 9.52 MHz/GPa[68]. One may distinguish two contributions to the variation of $D$ as a function of $P$: purely geometrical changes around the defect center and the variation of the defect's spin density. The former may be described using the 'compressed-orbital' model, introduced by Ivady *et al.*[68], according to which $D$ is scaled by a geometrical factor ($d/d_0$) determined by atomic relaxations under pressure, in proximity of the defect; $d$ and $d_0$ are neighbor distances under $P$ and at equilibrium, respectively. As shown in Fig. 5a, the compressed-orbital model describes well $D(P)$ in the case of diamond, showing a negligible contribution of spin density changes.

In contrast, $D(P)$ of the SiC divacancy is not well described by the compressed orbital model. As expected from the bulk modulus of SiC, which is substantially smaller than that of diamond, the divacancy defect structure relaxes significantly under pressure: $d/d_0$ ($P$) is 0.70 for $P$=100 GPa, compared to the value of 0.88 found for diamond NV under the same conditions. This relaxation allows for significant hybridization between the divacancy dangling bonds leading to large deviations of $D(P)$ from the values obtained with the compressed-orbital model. The slope of $D(P)$ close to ambient pressure is 16.34 MHz/GPa for the SiC divacancy.

Fig. 5b shows that the $D(P)$ of the Hf- and Zr-vacancy spins in 4$H$-SiC exhibits a behavior different from that reported in Fig.4: $D(P)$ deviates significantly from that predicted by a compressed orbital model, with a parabolic behavior and maxima around 70 GPa and 30 GPa for the Hf-vacancy and Zr-vacancy, respectively. In addition, close to $P$ = 0 GPa, the slope of $D(P)$ is about a factor of two smaller than that observed for the divacancy: 7.637 (2.835) MHz/GPa for the Hf-(Zr)vacancy. We note that the structural relaxation of the Hf- and the Zr-vacancy under pressure are relatively limited due to the presence of the LMIs, compared to the divacancy relaxation. At 100 GPa, $d/d_0$ is 0.84 for both Hf- and Zr-vacancy, to be compared to 0.7 of the SiC divacancy.

Fig. 5c shows $D(P)$ for Hf- and Zr-vacancies in $w$-AlN up to 30 GPa ($w$-AlN is known to undergo a structural phase transition above 20 GPa[69]). The figure indicates a greater sensitivity of $D(P)$ with respect to the corresponding defects in SiC, with slopes of 19.24 MHz/GPa and 15.03 MHz/GPa for the Hf- and the Zr-vacancy, respectively, in $w$-AlN. Our results show that the coupling characteristics of a defect spin qubit to lattice strain can vary over a wide range depending on its constituent electronic states (i.e. dangling bonds) and its host crystal as well.

Finally, as a guide for future EPR-based defect detections and to support development of the LMI-vacancy-based defects, we report computed hyperfine parameters ($A$) (see Eq. 1). The Hf- and Zr-vacancy defects may have intrinsic nuclear spins by implanting different isotopes: $^{177}$Hf (I=7/2, 18.6%), $^{179}$Hf (I=9/2, 13.62%), and $^{91}$Zr (I=5/2, 11.2%). The values of A are given in Table 6, for the $^{14}$N nuclear spin in diamond NV[70] and the LMI-vacancy defects. In 4$H$-SiC and $w$-AlN, there are also other intrinsic nuclear spins associated with $^{29}$Si (I=1/2, 4.7%), $^{13}$C (I=1/2, 1.1%), $^{27}$Al (I=5/2, 100%), and $^{14}$N (I=1, 99.63%). We report the hyperfine parameters for these intrinsic lattice nuclear spins coupled with the Hf-vacancy and the Zr-vacancy in 4$H$-SiC and $w$-AlN in Table S2 and S3, respectively.

## IV. DISCUSSIONS AND CONCLUSIONS

In this work, we proposed that large metal ion-vacancy complexes may be promising defect qubits in 4$H$-SiC and $w$-AlN. In particular, we considered Hf, Zr, La, and Y as they have larger ionic radii and smaller electronegativities than those of Si and Al. By using density functional theory, we showed that, similar to the diamond NV center and the SiC divacancy, the neutral Hf- and Zr-vacancy complexes are stable defects, with a $^3A_2$ spin-triplet ground state and an $^3E$ excited state, both with energies in the band gap of 4$H$-SiC and $w$-AlN. In addition, we found that the negatively charged La-vacancy and Y-vacancy complexes have a spin-triplet ground state, similar to the diamond NV center. However, in either 4$H$-SiC or $w$-AlN, the negative charge state of La and Y complexes is much shallower with respect to the CBM than the corresponding ones for the Hf- and Zr-vacancies. As a result, the $^3A_2$ - $^3E$ zero-phonon line excitation may ionize the La-vacancy defect center, making it unfavorable for use as optically addressable spin qubit. In order to guide future experiments, we calculated experimental observable of the Hf- and the Zr-vacancy in 4$H$-SiC and $w$-AlN, including optical zero-phonon lines, hyperfine parameters, and the zero-field splitting parameters.

Recently, Varley, Janotti, and Van de Walle also investigated impurity-vacancy complexes in *w*-AlN, including Ge, Sn, Ti, and Zr[36]. Using computational methods similar to those employed here, they suggested that Zr- and Ti-vacancy complexes would be good candidates for spin qubits in *w*-AlN. Their prediction on the Zr-vacancy is consistent with ours. In addition, Varley *et al.* have shown that the Ge-vacancy and the Sn-vacancy do not favor the S=1 state in *w*-AlN. We confirm this finding for 4*H*-SiC as well; our results show that in both crystals the S=1 state of the Ge-vacancy and the Sn-vacancy (see Fig. S7) is much higher in energy than their S=0 state, which is stabilized by charge transfer from neighboring dangling bonds.

The proposed LMI-vacancy defects may provide new opportunities to defect-based quantum technologies due to several unique features. For example, these defects may couple with various types of lattice strain[25,27-29,66]: we showed that the Hf-vacancy in *w*-AlN shows a large spin-pressure coupling which is about twice as large as that of the diamond NV, making it a good candidate for nano-scale pressure sensors[66]. Instead, the *D* parameter of the Zr-vacancy in 4*H*-SiC showed the smallest sensitivity to pressure, which may be useful in applications requiring spin sub-level structure insensitive to pressure. Work is in progress to explore spin responses to uniaxial strains, which may be useful in applications ranging from nano-scale sensing[71] to creation of hybrid quantum systems[30-32].

Nuclear spins associated with different isotopes of Hf and Zr ($^{177}$Hf (I=7/2, 18.60%), $^{179}$Hf (I=9/2, 13.62%), $^{91}$Zr (I=5/2, 11.22%)) may also be used as quantum resources[72]. For example, Klimov *et al.*, demonstrated a coherent coupling between a divacancy-related (PL5) spin and native nuclear spins associated with $^{13}$C and $^{29}$Si isotopes at room temperature[73]. This study was a milestone towards developing SiC-based hybrid quantum systems. However, it is still challenging to find $^{29}$Si and $^{13}$C nuclear spins strongly coupled to a divacancy spin due to their natural abundances: 4.7% for $^{29}$Si and 1.1% for $^{13}$C. The nuclear spins of Hf and Zr may resolve this issue and provide intrinsic nuclear spins at a well-defined position of the LMI-vacancy complexes. Finally, the use of LMIs may be beneficial for defect localization. For example, in the case of divacancy or Si vacancy in SiC, it is hard to control the position of the defects as both C vacancies and Si vacancies are highly mobile. The mobility of Hf and Zr in SiC would be much lower than that of the C vacancy and the Si vacancy due to their large mass.

In summary, optically addressable spins bound to point defects in solids have a great potential for quantum information processing, quantum communications, and hybrid quantum systems. The defect complexes proposed here would provide alternative quantum systems in heterogeneous materials such as 4*H*-SiC and *w*-AlN that could broaden the scope of defect-based quantum technologies.

**ACKNOWLEDGEMENTS**

We thank William Koehl, Sam Whiteley, Chris Anderson, Gary Wolfowicz, Joseph Heremans, and David Awschalom for helpful discussions. We are grateful to Viktor Ivády for his help in the numerical implementation of zero-field splitting calculations. HS is supported by the National Science Foundation (NSF) through the University of Chicago MRSEC under award number DMR-1420709 and by ANL LDRD. GG, MG and HM are supported by MICCoM, as part of the Computational Materials Sciences Program funded by the U.S. Department of Energy, Office of Science, Basic Energy Sciences, Materials Sciences and Engineering Division. This research used resources of the National Energy Research Scientific Computing Center (NERSC), a DOE Office of Science User Facility supported by the Office of Science of the U.S. Department of Energy under Contract No. DE-AC02-05CH11231, resources of the Argonne Leadership Computing Facility, which is a DOE Office of Science User Facility supported under Contract DE-AC02-06CH11357, and resources of the University of Chicago Research Computing Center.

**FIGURES**

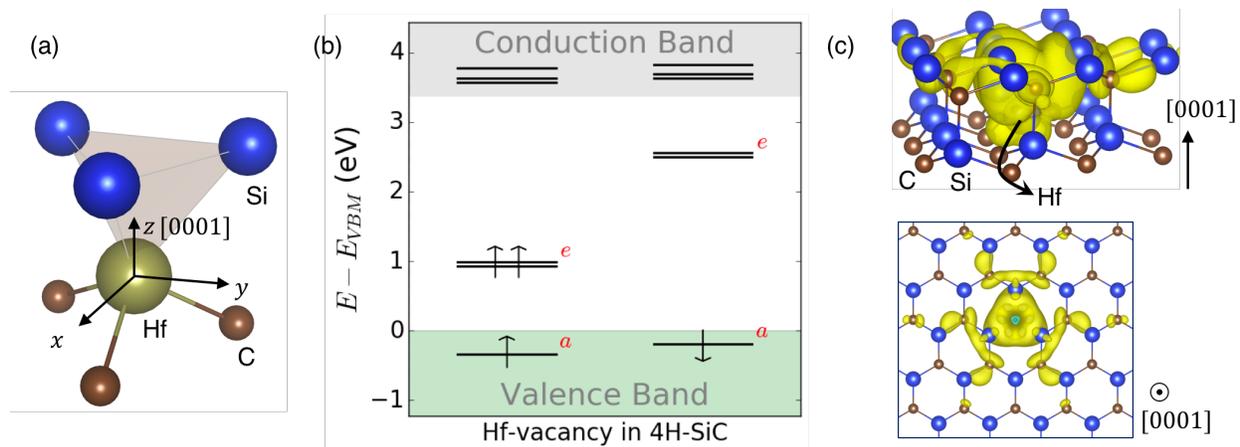

**Figure 1. Hf-vacancy complex in 4*H*-SiC.** (**a**) Proposed defect structure of a Hf-vacancy complex in 4*H*-SiC with (*hh*) axial configuration and $C_{3v}$ symmetry: Hf substitutes Si at an *h*-site and it pairs with a C vacancy at an *h*-site. Only the nearest neighboring Si and C atoms are shown for clarity. (**b**) The defect level diagram of the Hf-vacancy complex calculated at the DFT- DDH hybrid level of theory. The totally symmetric *a* state is located at -0.34 eV and -0.19 eV below the valence band edge in the spin-up and the spin-down channel, respectively. (**c**) Side (up) and top (down) views of the ground-state spin density of the Hf-vacancy defect calculated at the DDH level of theory.

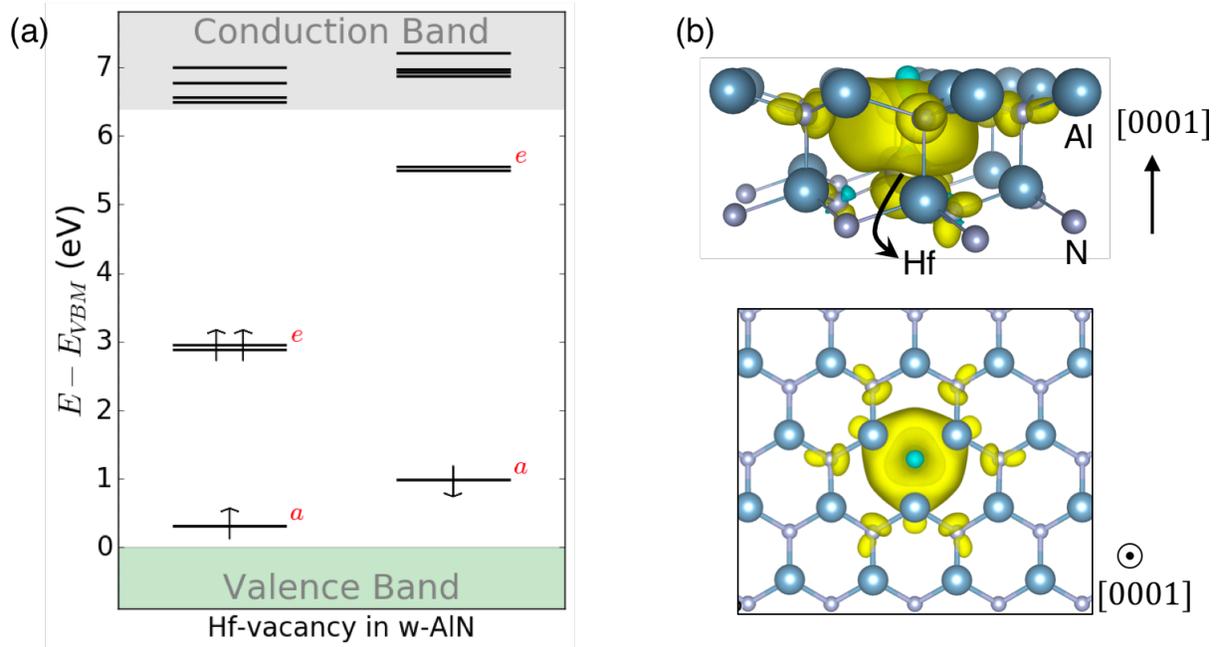

**Figure 2. Hf-vacancy complex in *w*-AlN.** (**a**) The defect level diagram of an axial Hf-vacancy in *w*-AlN calculated at the DDH level of theory. The symmetry of the state is $^3A_2$. In this study, we only consider the axial defect configuration in $C_{3v}$ symmetry. In principle, however, a basal configuration in $C_{1h}$ symmetry is also possible. (**b**) Side (top) and top (bottom) views of the ground-state spin density of the Hf-vacancy in *w*-AlN calculated at the DDH level of theory.

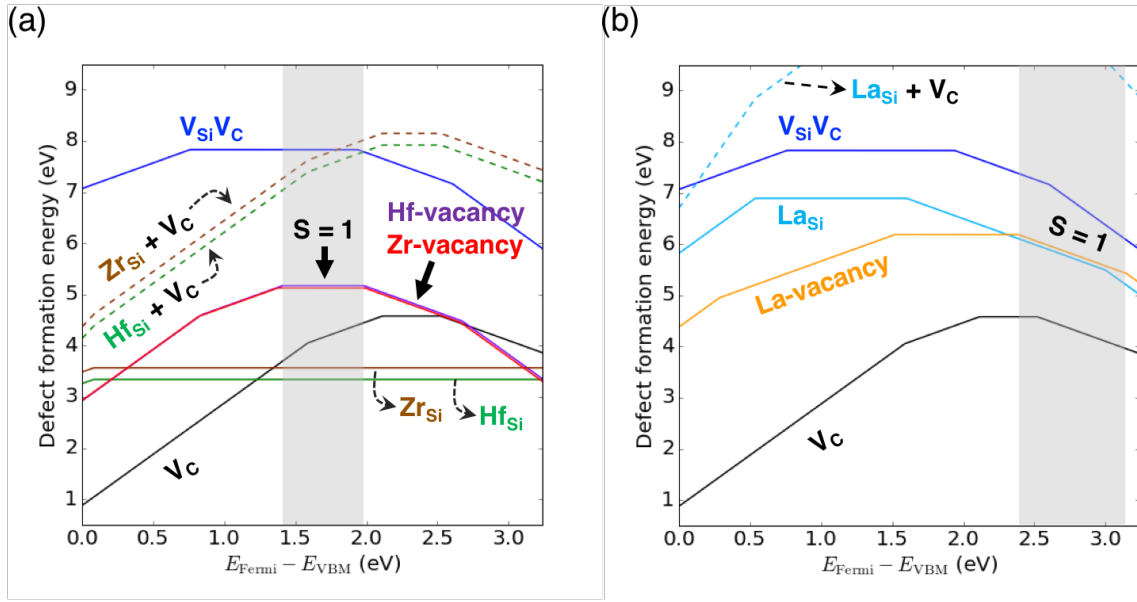

**Figure 3. Defect formation energy of spin defects in 4*H*-SiC. (a,b)** Defect formation energy of Hf- and Zr-related defects in 4*H*-SiC (a), and that of La-related defects in 4*H*-SiC (b) as a function of Fermi level referred to the valence band maximum (VBM). Calculations were conducted at the DFT-DDH level of theory (see text). The defect formation energy of the (*hh*)-divacancy is included for comparison. For simplicity, the results of Y-related defects are reported in Fig. S6. The dotted lines are the sum of the formation energies of substitutional impurity (either $Hf_{Si}$, $Zr_{Si}$, or $La_{Si}$) and C vacancy to be compared to that of the corresponding LMI-vacancy defect complex. The grey shaded area in each plot indicates a Fermi-level range, in which the LMI-vacancy complexes exhibit a stable $^3A_2$ spin-triplet (S=1) ground state in 4*H*-SiC.

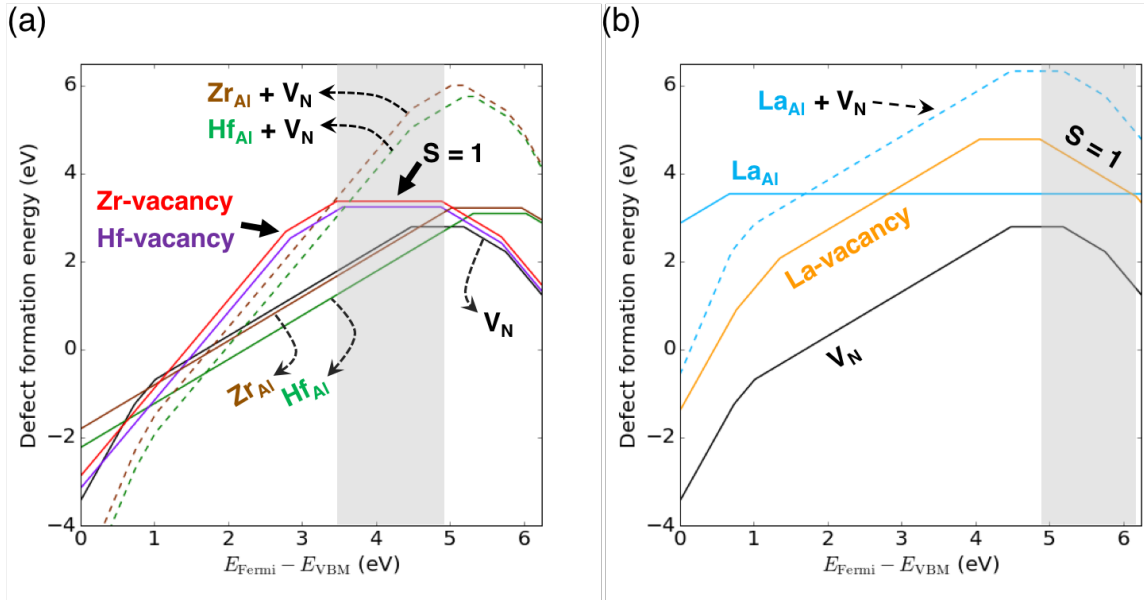

**Figure 4. Defect formation energy of spin defects in *w*-AlN. (a,b)** Defect formation energy of Hf- and Zr-related defects in *w*-AlN (a), and that of La-related defects in *w*-AlN (b) The DDH-DFT was used. The formation energy of N vacancy, which is a common defect in *w*-AlN, is included for comparison. The dotted lines are the sum of the formation energies of a substitutional impurity (either $Hf_{Al}$ or $Zr_{Al}$) and a N vacancy to be compared to that of the corresponding LMI-vacancy defect complex. The grey shaded area in each plot indicates a Fermi-level range where the LMI-vacancy complexes have stable $^3A_2$ spin-triplet (S=1) ground state in *w*-AlN.

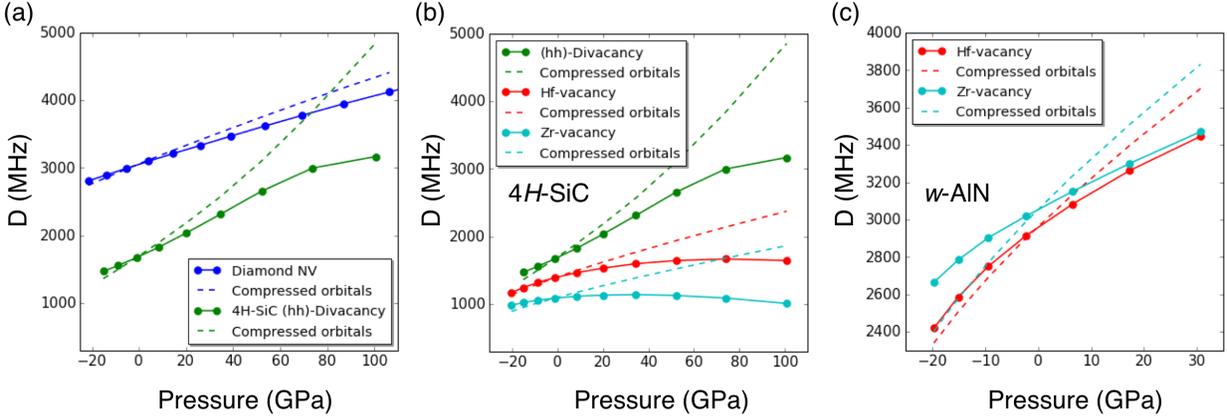

**Figure 5. Zero-field splitting (ZFS) of the spin defects in 4*H*-SiC and *w*-AlN. (a)** ZFS parameters (D) of the diamond NV and the SiC divacancy as a function of hydrostatic pressure. **(b, c)** ZFS parameters (D) of the Hf-vacancy and the Zr-vacancy as a function of hydrostatic pressure in 4*H*-SiC (b) and in *w*-AlN (c). For the defects in 4*H*-SiC, we also show D of the divacancy for comparison. We considered a pressure range from -20 GPa to 100 GPa, in which 4*H*-SiC is known to be stable[67]. For defects in w-AlN under pressure, we considered a pressure range from -20 to to 30 GPa as *w*-AlN is known to undergoes a structural phase transition above 20 ~ 30 GPa[69].

## TABLES

**Table 1.** Computed bulk properties of the 4*H*-SiC calculated at the PBE and the DDH-DFT levels of theory along using ONCV pseudopotentials[41,42]. Experimental values are from Ref.[40,47]

|  | Lattice parameters | | Dielectric constants | |
| --- | --- | --- | --- | --- |
|  | $a$ (Å) | $c$ (Å) | Electronic $(\epsilon_{\infty,\parallel}/\epsilon_{\infty,\perp})$ | Static $(\epsilon_{0,\parallel}/\epsilon_{0,\perp})$ |
| PBE | 3.096 | 10.136 | 6.938 / 7.251 | 10.306 / 10.938 |
| DDH | 3.087 | 10.089 | 6.396 / 6.623 | 9.663 / 9.926 |
| Experiment | 3.073 | 10.053 | 6.52 / 6.70 | 9.66 / 10.03 |

**Table 2.** Computed band-gaps (eV) of the crystals considered in this study calculated at the $G_0W_0$@PBE, the DDH hybrid, and the HSE06 hybrid functional levels of theory.

| Host crystals | DD-hybrid (eV) | HSE06 (eV) | $G_0W_0$@PBE (eV) | Experiment |
|---|---|---|---|---|
| Diamond | 5.59 | 5.42 | 4.25 | 5.48[74] |
| 4H-SiC | 3.28 | 3.19 | 3.29 | 3.23[40] |
| w-AlN | 6.39 | 5.67 | 6.12 | 6.03 - 6.28[75] |

**Table 3.** Computed Zero-field splitting parameters (*D*) of the diamond NV center, the divacancy spins in 4H-SiC, and the Hf- and Zr-vacancy complexes in 4H-SiC and w-AlN. The single-particle wavefunctions for the defects were calculated using the Quantum Espresso code with the ONCV[41,42] and the PAW[49] pseudopotentials.

| Host crystals | Defects | Theory (GHz) (This work) (QE + ONCV) | Theory (GHz) (This work) (QE + PAW) | Theory (GHz) (Previous work[28]) (VASP + PAW) | Exp.[58,59] (GHz) |
|---|---|---|---|---|---|
| Diamond | NV center | 3.03 | 2.90 | 2.854 | 2.88 |
| 4H-SiC | (*hh*)-divacancy | 1.682 | 1.387 | 1.358 | 1.336 |
| | (*hk*)-divacancy | 1.580 | 1.306 | 1.320 | 1.222 |
| | (*kh*)-divacancy | 1.641 | 1.356 | 1.376 | 1.334 |
| | (*kk*)-divacancy | 1.635 | 1.349 | 1.321 | 1.305 |
| | Hf-vacancy | 1.403 | 1.291 | n/a | n/a |
| | Zr-vacancy | 1.096 | 1.035 | n/a | n/a |
| w-AlN | Hf-vacancy | 2.962 | 2.896 | n/a | n/a |
| | Zr-vacancy | 3.053 | 2.925 | n/a | n/a |

**Table 4.** Computed energy levels (eV) of the occupied spin-up (left number) and unoccupied spin-down (right number) *e*-manifolds of the LMI-vacancy complexes in 4*H*-SiC and *w*-AlN with respect to the valence band edge using the $G_0W_0$@PBE, the DDH functional, and the HSE06 hybrid functional levels of theory. The experimental band gap ($E_g$) of the materials are given. The computed band gaps are reported in Table 2.

| Host crystals | Defects | $G_0W_0$ (eV) | DD-hybrid (eV) | HSE06 (eV) |
|---|---|---|---|---|
| 4*H*-SiC | Hf-vacancy | 0.97 / 2.26 | 0.96 / 2.54 | 0.99 / 2.48 |
| ($E_g$ = 3.3 eV) | Zr-vacancy | 1.05 / 2.35 | 0.93 / 2.54 | 0.97 / 2.50 |
| *w*-AlN | Hf-vacancy | 2.92 / 4.96 | 2.92 / 5.53 | 2.90 / 4.78 |
| ($E_g$ = 6.2 eV) | Zr-vacancy | 3.01 / 5.12 | 2.83 / 5.56 | 2.82 / 4.82 |

**Table 5.** Computed zero-phonon lines (eV) of the (*hh*)-divacancy and the LMI-vacancy complexes (Hf and Zr only) in 4*H*-SiC and *w*-AlN using various levels of theory; the semi-local PBE functional, the DDH functional, and the HSE06 hybrid functional. Spin-conserving intra-defect excitation between the $^3A_2$ ground state and the $^3E$ excited state was considered.

| Host crystals | Defects | PBE (eV) | DD-hybrid (eV) | HSE06 (eV) | Experiment (eV) |
|---|---|---|---|---|---|
| Diamond | NV center | 1.72 | 2.22 | 2.23 | 1.945[5] |
| 4*H*-SiC | (*hh*)-divacancy | 1.03 | 1.30 | 1.33 | 1.094[9] |
| | Hf-vacancy | n/a | 2.04 | 2.13 | n/a |
| | Zr-vacancy | n/a | 1.96 | 2.05 | n/a |
| *w*-AlN | Hf-vacancy | 2.46 | 3.07 | 2.88 | n/a |
| | Zr-vacancy | 2.33 | 2.98 | 2.79 | n/a |

**Table 6.** Computed hyperfine parameters (MHz) for the Hf-vacancy and Zr-vacancy complexes in 4$H$-SiC and $w$-AlN. For comparison, the computed hyperfine parameters of the diamond NV center are also reported along with the experimental data[70] in parenthesis. Other hyperfine parameters are reported in Table S2 and S3.

| Host crystals | Defects | Nuclear spin | $A_{xx}$ (MHz) | $A_{yy}$ (MHz) | $A_{zz}$ (MHz) |
|---|---|---|---|---|---|
| Diamond | NV center | $^{14}$N (I=1, 99.6%) | -2.02 (-2.14) | -2.15 (-2.70) | -2.02 (-2.14) |
| 4$H$-SiC | Hf-vacancy | $^{177}$Hf (I=7/2, 18.6%) | 7.58 | 7.91 | -8.60 |
|  |  | $^{179}$Hf (I=9/2, 13.62%) | -4.76 | -4.97 | 5.40 |
|  | Zr-vacancy | $^{91}$Zr (I=5/2, 11.2%) | 1.92 | 1.70 | 17.57 |
| $w$-AlN | Hf-vacancy | $^{177}$Hf (I=7/2, 18.6%) | 26.05 | 26.20 | 10.53 |
|  |  | $^{179}$Hf (I=9/2, 13.62%) | -16.36 | -16.46 | -6.62 |
|  | Zr-vacancy | $^{91}$Zr (I=5/2, 11.2%) | 6.21 | 6.12 | 15.59 |

# Supplementary Information for:
# Designing defect-based qubit candidates in wide-gap binary semiconductors for solid-state quantum technologies

Hosung Seo[1,2,3], He Ma[2,4], Marco Govoni[1,2], and Giulia Galli[1,2]*

5. Materials Science Division, Argonne National Laboratory, Argonne, IL 60439, USA
6. The Institute for Molecular Engineering, The University of Chicago, Chicago, IL 60615, USA
7. Department of Physics, Ajou University, Suwon, Republic of Korea
8. Chemistry Department, The University of Chicago, Chicago, IL 60615, USA

## 1. GW convergence test

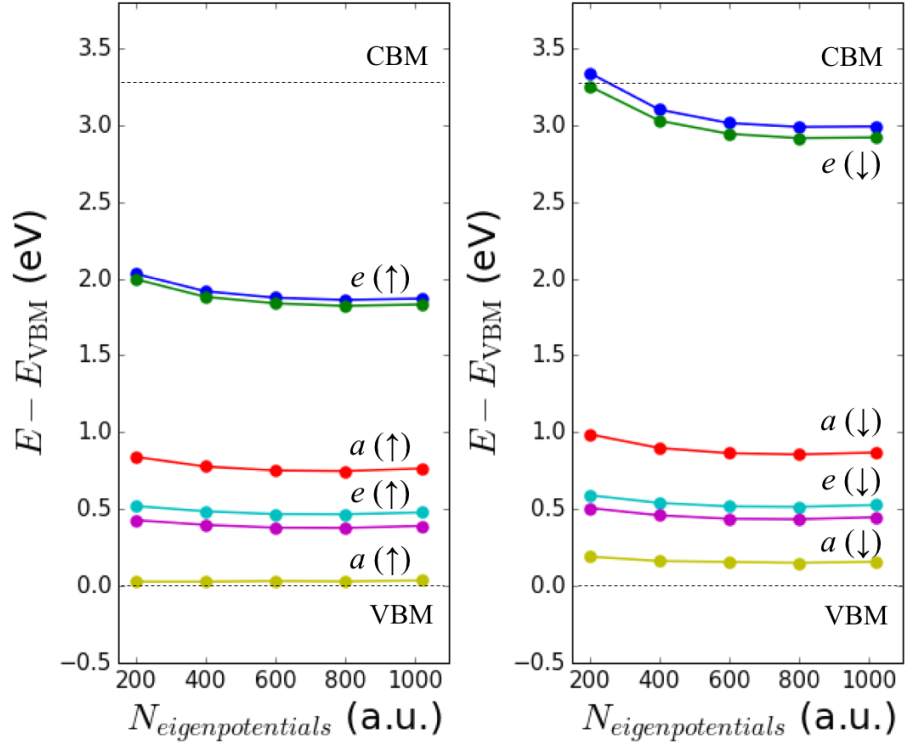

**Figure S1. Convergence test for GW calculations of defects in SiC.** Defect levels with respect to the valence band edge of the negatively charged La-vacancy in 4*H*-SiC are calculated as a function of the number of eigenpotentials used in the GW calculation. The left panel and the right panel show the defect levels in the spin-up and the spin-down channels, respectively. We found that 1024 eigenpotentials are enough to converge the quasi-particle defect levels within 0.01 eV.

## 2. First-principles calculations of the zero-field splitting

To calculate the zero-field splitting parameters of the LMI-vacancy spins, we used the wavefunctions obtained with the norm-conserving ONCV pseudopotentials. However, as shown in Table 3, if we switch the pseudopotentials to PAW pseudopotentials instead, then we find our results are in excellent agreement with the experimental results and the previous theoretical results[1] obtained using PAW pseudopotentials. However, the agreement between the PAW results (both ours and the previous calculations[1]) and the experimental values may not be reliable. In both our and previous PAW calculations of the D parameters, only the pseudo-wavefunctions were used while the core-region electrons were not treated properly. Therefore, the agreement between the PAW results and the experimental results would be due to some error cancellation. Therefore, we decided to mainly use our ONCV PP results, in which the core region electrons are at least properly normalized. To resolve the discrepancy between the theoretical values and the experimental results, further studies might be useful, for example, using different density functionals.

## 3. The electronic structure of the diamond NV, the SiC divacancy, and the LMI-vacancy complexes in 4*H*-SiC and *w*-AlN

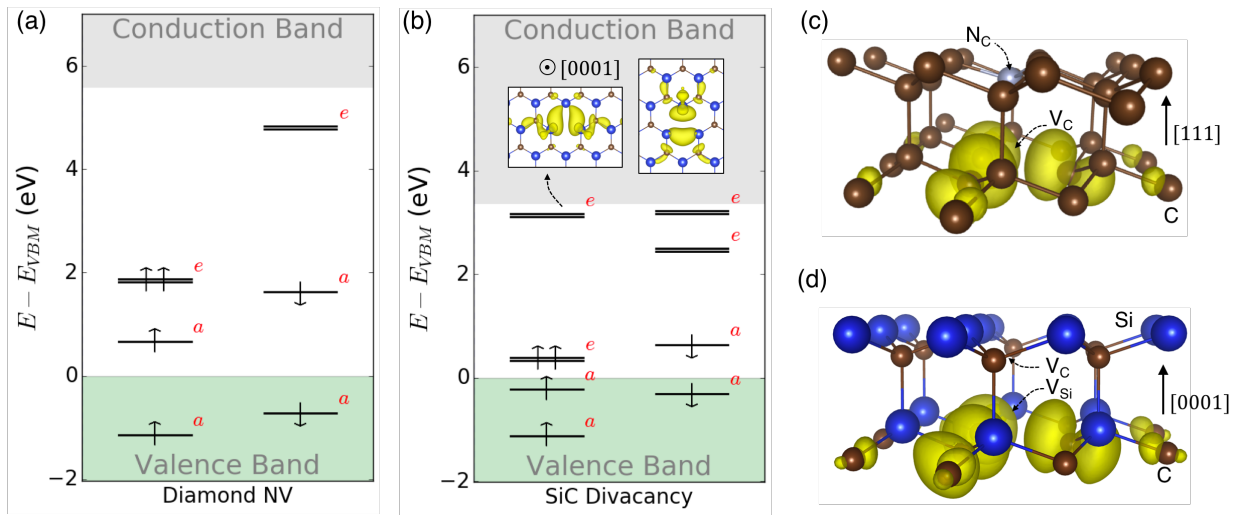

**Figure S2. Diamond NV and SiC divacancy. (a,b)** The defect level diagrams of the diamond NV (a) and the (*hh*)-divacancy in 4*H*-SiC (b) calculated using the DDH-DFT. In both cases, the occupied $e_x$ and $e_y$ states are mainly derived from the C $2sp^3$ dangling bonds and dominantly contribute to the ground-state spin density. Unlike the diamond NV, the SiC divacancy has another e manifold that is also formed inside the band gap of 4*H*-SiC. The empty $e_x$ and $e_y$ states are mainly composed of the Si $3sp^3$ dangling bonds and they also localize in space as shown by their electron charge density. **(c,d)** The ground-state spin densities of the diamond NV (a) and the (*hh*)-divacancy in 4*H*-SiC (b). Both defects have the same $C_{3v}$ symmetry. The $C_{3v}$ axis for the diamond NV and the divacancy is parallel to the (111) direction of the diamond lattice and the (0001) direction of the 4*H*-SiC lattice, respectively.

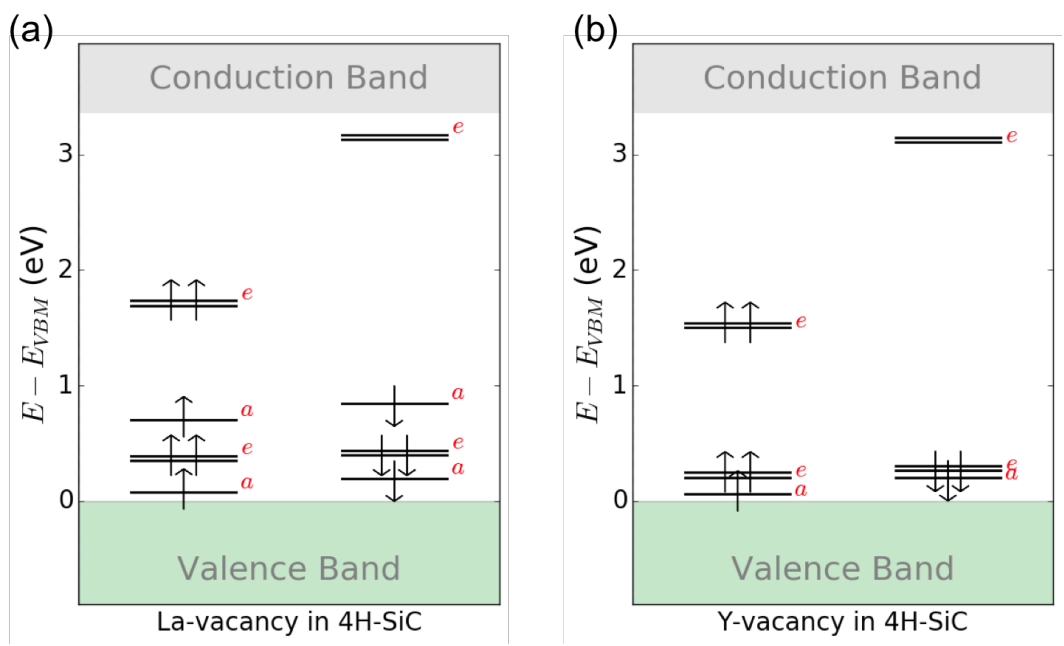

**Figure S3.** Defect level diagram of negatively charged La-vacancy (a) and Y-vacancy (b) in 4*H*-SiC calculated using the dielectric dependent hybrid functional theory. The valence band maximum (VBM) is set to be 0 eV.

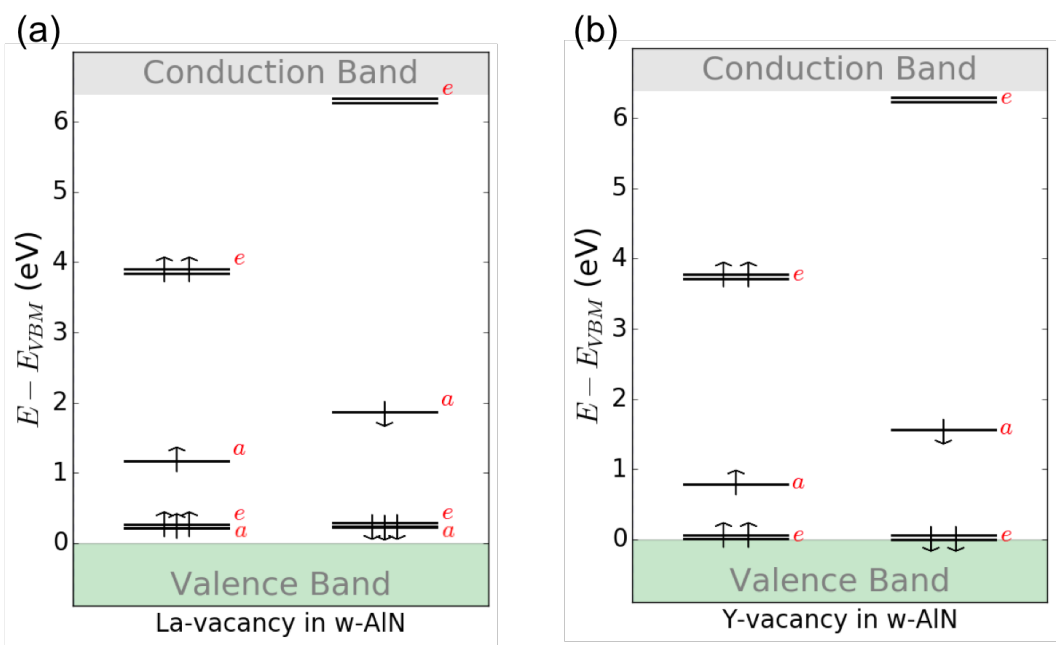

**Figure S4.** Defect level diagram of negatively charged La-vacancy (a) and Y-vacancy (b) in *w*-AlN calculated using the dielectric dependent hybrid functional theory. The valence band maximum (VBM) is set to be 0 eV.

## 4. Defect formation energy

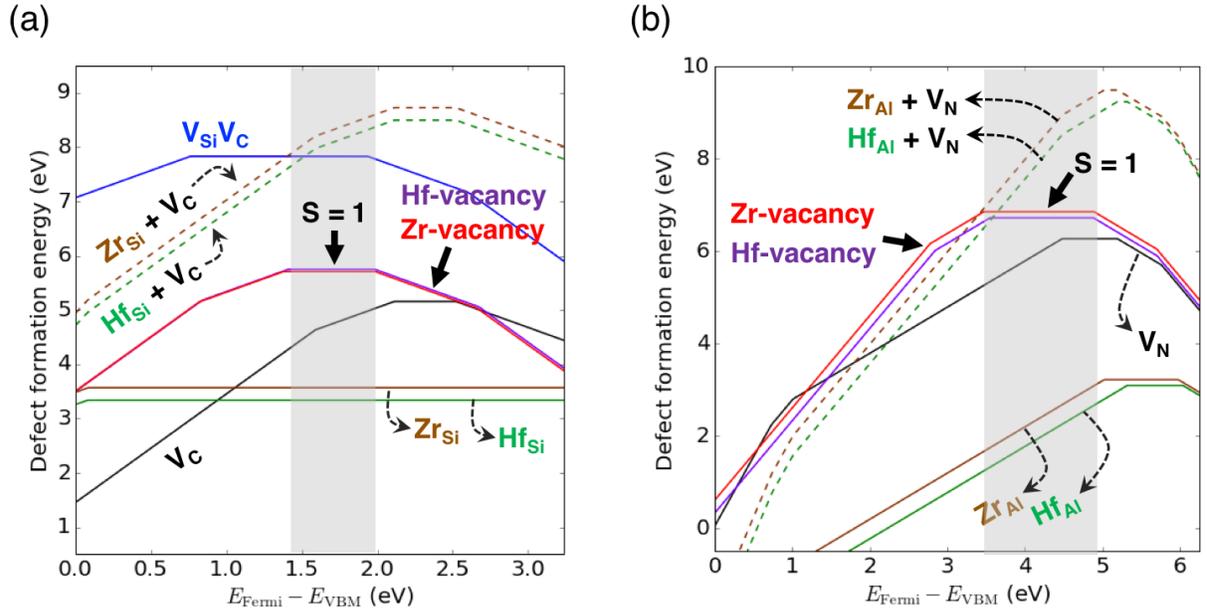

**Figure S5.** Defect formation energy of Hf- and Zr-related defects in 4*H*-SiC in the C-rich limit (a) and those in *w*-AlN in N-rich limit (b) as a function of Fermi level with respect to the valence band maximum (VBM). The dielectric-dependent hybrid density functional theory was employed. A dotted line is a sum of the formation energies of substitutional impurity (either $Hf_{Si}$ and $Zr_{Si}$) and anion vacancy (C vacancy for SiC and N vacancy for AlN) to be compared to that of the corresponding LMI-vacancy defect complex. The grey shaded area in each plot indicates a Fermi-level range where the LMI-vacancy complexes have stable $^3A_2$ spin-triplet (S=1) ground state in 4*H*-SiC.

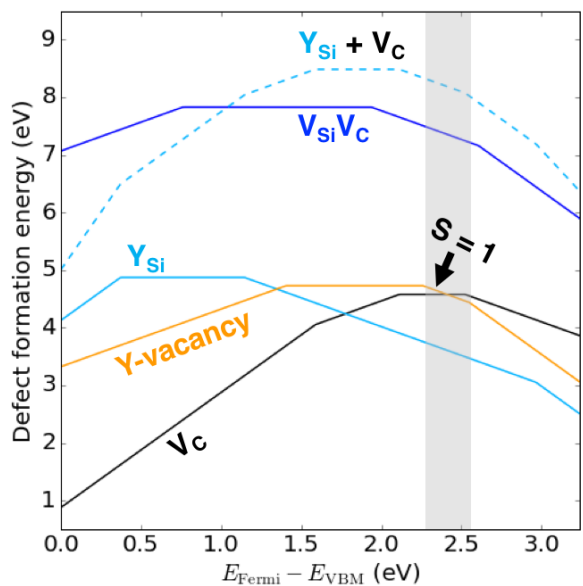

**Figure S6.** Defect formation Energy of Y-related defects in 4*H*-SiC in the C-poor limit as a function of Fermi level with respect to the valence band maximum (VBM). The dielectric-dependent hybrid density functional theory was employed. The defect formation energy of the more well-established divacancy defect is included for comparison. The dotted line is a sum of the formation energies of substitutional Y impurity and C vacancy to be compared to that of the corresponding Y-vacancy defect complex. The grey shaded area in each plot indicates a Fermi-level range where the Y-vacancy complex has stable $^3A_2$ spin-triplet (S=1) ground state in 4*H*-SiC.

## 5. Defect level diagram of Ge- and Sn-vacancy in *w*-AlN

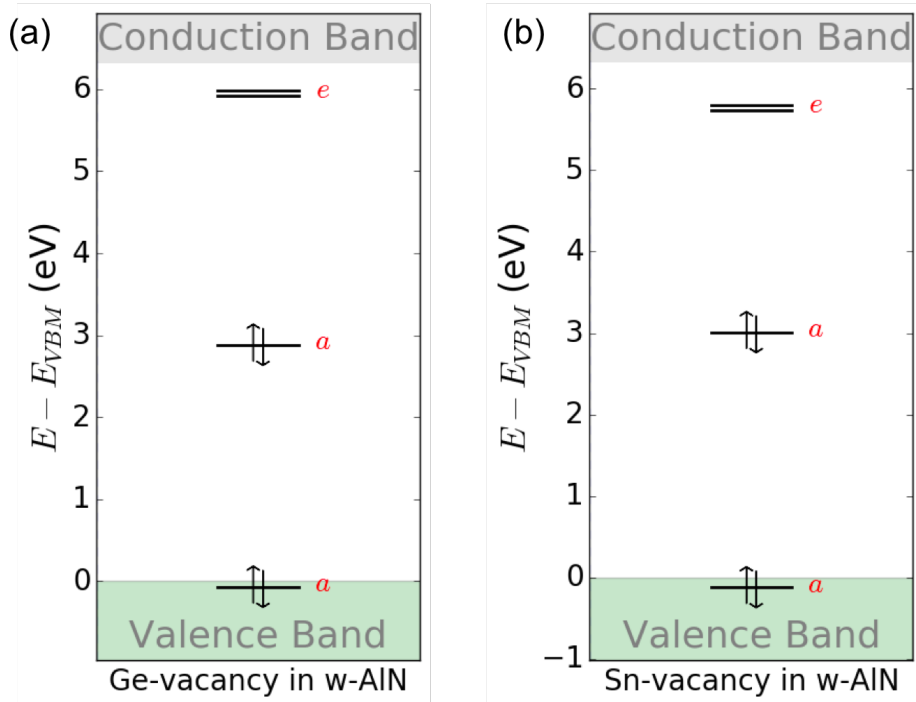

**Supplementary Figure 7.** Defect level diagrams of Ge-vacancy (a) and Sn-vacancy (b) in w-AlN calculated using the dielectric-dependent hybrid functional theory. The ground state of the both defects is spin-singlet (S=0).

## 6. Energetic stability of the S=1 state of the LMI-vacancy complexes

**Table S1.** Computed total energy differences (meV) between the spin-triplet ($S=1$) state and the spin-singlet state (S=0) of the LMI-vacancy complexes in 4*H*-SiC and *w*-AlN. The DDH functional was used along with the 240-atom supercell and a k-point mesh of 2×1×1. A negative energy difference means that the $S=1$ solution is lower in energy. For Hf-vacancy and Zr-vacancy, the neutral state ($q = 0$) was considered, while the negative charge state ($q = -1$) was considered for La-vacancy and Y-vacancy.

|  | Hf-vac. ($q = 0$) | Zr-vac. ($q = 0$) | La-vac. ($q = -1$) | Y-vac. ($q = -1$) |
|---|---|---|---|---|
|  | (meV) | (meV) | (meV) | (meV) |
| 4*H*-SiC | -205 | -240 | -127 | -173 |
| *w*-AlN | -380 | -438 | -209 | -205 |

## 7. Hyperfine coupling parameters

**Table S2.** Hyperfine parameters of spin defects in 4*H*-SiC. For comparison, the computed hyperfine parameters of the (*hh*)-divacancy in 4*H*-SiC are also reported along with the experimental data[2] in parenthesis. $Si_I$ and $C_I$ atoms are the nearest neighboring silicon and carbon atoms of a given defect. For Hf-vacancy complex, for example, $C_I$ atoms are atoms bonded to the Hf impurity and $Si_I$ atoms are the nearest neighboring Si atoms of the C vacancy site.

| Defects | Nuclear spin | $A_{xx}$ (MHz) | $A_{yy}$ (MHz) | $A_{zz}$ (MHz) |
|---|---|---|---|---|
| (*hh*)-Divacancy | $^{29}Si_I$ (I=1/2, 4.7%) | -1.34 (3) | -1.20 (4) | -2.59 (5) |
|  | $^{13}C_I$ (I=1/2, 1.1%) | 58.74 (56) | 58.00 (55) | 122.76 (116) |
| Hf-vacancy | $^{29}Si_I$ (I=1/2, 4.7%) | -98.96 | -98.20 | -138.84 |
|  | $^{13}C_I$ (I=1/2, 1.1%) | -0.25 | -0.22 | -0.76 |
| Zr-vacancy | $^{29}Si_I$ (I=1/2, 4.7%) | -88.34 | -87.58 | -125.59 |
|  | $^{13}C_I$ (I=1/2, 1.1%) | 0.22 | 0.32 | -0.40 |

**Table S3.** Hyperfine parameters of spin defects in *w*-AlN. $Al_I$ and $N_I$ atoms are the nearest neighboring Al and N atoms of a given defect. For Hf-vacancy complex, for example, $N_I$ atoms are atoms bonded to the Hf impurity and $Al_I$ atoms are the nearest neighboring Al atoms of the N vacancy site.

| Defects | Nuclear spin | $A_{xx}$ (MHz) | $A_{yy}$ (MHz) | $A_{zz}$ (MHz) |
|---|---|---|---|---|
| Hf-vacancy | $Al_I$ | 138.40 | 136.98 | 161.12 |
|  | $N_I$ | -0.49 | -0.82 | 0.86 |
| Zr-vacancy | $Al_I$ | 121.31 | 120.15 | 142.55 |
|  | $N_I$ | -0.53 | -1.03 | 0.85 |

## Supplementary Reference